\documentclass[aip,rsi,amsmath,amssymb,preprint]{revtex4-1}

\usepackage{graphicx}
\usepackage{dcolumn}
\usepackage{bm}
\usepackage{subcaption}

\usepackage[english]{babel}

\begin{document}

\title{Compact ultracold electron source based on a grating magneto-optical trap}

\author{J.G.H. Franssen}
\affiliation{Department of Applied Physics, Eindhoven University of Technology, P.O. Box 513, 5600 MB Eindhoven, The Netherlands}
\affiliation{Institute for Complex Molecular Systems, Eindhoven University of Technology, P.O. Box 513, 5600 MB Eindhoven, The Netherlands}
\author{T.C.H. de Raadt}
\affiliation{Department of Applied Physics, Eindhoven University of Technology, P.O. Box 513, 5600 MB Eindhoven, The Netherlands}
\author{M.A.W. van Ninhuijs}
\affiliation{Department of Applied Physics, Eindhoven University of Technology, P.O. Box 513, 5600 MB Eindhoven, The Netherlands}
\author{O.J. Luiten}
\email{o.j.luiten@tue.nl}
\affiliation{Department of Applied Physics, Eindhoven University of Technology, P.O. Box 513, 5600 MB Eindhoven, The Netherlands}
\affiliation{Institute for Complex Molecular Systems, Eindhoven University of Technology, P.O. Box 513, 5600 MB Eindhoven, The Netherlands}

\date{\today}

\begin{abstract}

The ultrafast and ultracold electron source, based on near-threshold photoionisation of a laser-cooled and trapped atomic gas, offers a unique combination of low transverse beam emittance and high bunch charge. Its use is however still limited because of the required cold-atom laser-cooling techniques. Here we present a compact ultracold electron source based on a grating magneto-optical trap (GMOT), which only requires one trapping laser beam that passes through a transparent accelerator module. This makes the technique more widely accessible and increases its applicability. We show the GMOT can be operated with a hole in the center of the grating and with large electric fields applied across the trapping region, which is required for extracting electron bunches. The calculated values of the applied electric field were found to agree well with measured Stark shifts of the laser cooling transition. The electron beams extracted from the GMOT have been characterised. Beam energies up to $10~\rm{keV}$ were measured using a time-of-flight method. The normalised root-mean-squared transverse beam emittance was determined using a waist scan method, resulting in $\epsilon=1.9~\rm{nm}\cdot\rm{rad}$. The root-mean-squared transverse size of the ionisation volume is $30~\mu\rm{m}$ or larger, implying an electron source temperature in the few-$10\rm{K}$ range, $2-3$ orders of magnitude lower than conventional electron sources, based on photoemission or thermionic emission from solid state surfaces.

\end{abstract}


\maketitle

\section{Introduction}

In the past decade new tools have emerged that allow investigation of structural dynamics with atomic spatial and temporal resolution, i.e. as small as $0.1~\rm{nm}$ and $100~\rm{fs}$: Ultrafast Electron Microscopy\cite{Zewail1,Zewail2,Baum2007,king2005} (UEM), Ultrafast Electron Diffraction\cite{firstUEDRF,Siwick2003,Musumeci20101,RFUED2} (UED) and X-ray crystallography using Free Electron Lasers (XFELs)\cite{chapman2011,Boutet1217737,Kang2015}. This revolution would not have been possible without the spectacular development of ultrafast pulsed electron sources\cite{RFGUN5}. By femtosecond photoemission from flat photocathodes in RF photoguns, highly charged electron bunches can be created of sufficient quality to drive XFELs, enabling single-shot, femtosecond X-ray diffraction of protein nanocrystals~\cite{chapman2011}. Unfortunately XFELs are big and costly facilities with limited access for the average researcher. In an alternative approach the electron bunches that drive the XFELs can also be directly used for single-shot UED~\cite{firstUEDRF,RFUED1,RFUED2}.

Using electrons instead of X-rays has the great advantage of smaller, cheaper setups. However, the beam quality is not sufficient for studying complicated macromolecular structures or for imaging with (sub)nanometer resolution. Higher beam quality is generally provided by sharp-tipped sources, developed for electron microscopy. By sideways femtosecond photoemission from a nanometer-sized field emission tip~\cite{RopersSharpTip} (or by RF chopping~\cite{Oldfield,VERHOEVEN201885,VANRENS201877}) the same beam quality can be achieved as in conventional electron microscopy, enabling imaging with atomic spatial and temporal resolution. However, this results on average in less than one electron per pulse. Increasing the charge per pulse spoils the beam quality and therefore the atomic resolution. A source that offers the combination of high beam quality and high bunch charge is highly desirable. 

In the quest for better beam coherence while maintaining a large source size, a new electron source was proposed~\cite{proposalUCES}, the ultracold electron source (UCES). In the UCES the initial transverse angular momentum spread is decreased which results in increased beam coherence for a given source size~\cite{Engelen2013,Engelen2014,McCulloch2013}. This significantly reduces Coulomb effects at the source which allows extraction of more charge, required for single-shot measurements.

In the UCES high charge electron bunches are created by near threshold photo-ionisation of a cloud of laser-cooled and trapped atoms in a magneto optical trap (MOT)~\cite{metcalf}. Previous work showed high quality diffraction patterns~\cite{VanMourik2014a,Speirs2015a} with these bunches, demonstrating pulsed electron source temperatures as low~\cite{Engelen2013,McCulloch2013,Engelen2014} as a few-$10~\rm{K}$. Additionally it was shown that it is possible to extract ultracold picosecond electron pulses~\cite{franssen_pulse} which can in principle be compressed to $\sim 100\rm{fs}$ using well established RF compression techniques~\cite{VanOudheusden2010a}. 

In this work we present a novel compact ultracold electron source based on a grating magneto-optical trap (GMOT), which only requires one trapping laser beam. This makes the technique more widely accessible and therefore increases its applicability. The paper is organised as follows: In Section~\ref{secGmot} we describe the operating principle of a GMOT. Next, in Section~\ref{design}, we discuss the design of the ultracold electron source, the vacuum system, the accelerator and the electron beamline. In Section~\ref{sectionexperi} we will show that it is possible to operate a GMOT with a hole in the center of the grating and with large electric fields applied across the trapping region, which are both required for extracting electron bunches. We will also show that the calculated values of the applied electric field agree well with measured Stark shifts of the laser cooling transition. Finally, in section~\ref{ucescomm} we will discuss the commissioning of the ultracold electron source. The electron beams extracted from the GMOT have been characterised. Beam energies up to $10~\rm{keV}$ were measured using a time-of-flight method. The normalised root-mean-squared (rms) transverse beam emittance $\epsilon$ was determined using a waist scan method, resulting in $\epsilon=1.9~\rm{nm}\cdot\rm{rad}$. The rms transverse source size is $30~\mu\rm{m}$ or larger, implying an electron source temperature less than $25~\rm{K}$.

\section{The principle of a grating MOT based UCES}\label{secGmot}

In a conventional magneto-optical trap (MOT) atoms are laser cooled using three pairs of orthogonal laser beams whose frequency is red shifted with respect to the atomic laser cooling transition. The atoms are trapped using a quadrupole magnetic field which creates position-dependent resonance conditions\cite{metcalf} for the atomic transition through the Zeeman effect, and thus a restoring force and stable trapping. A recent development in the field of laser cooling and trapping is the Grating MOT\cite{nshii2013} which requires only one input laser beam instead of the six for a conventional MOT.

\subsection{Grating MOT}\label{MOTvol}

\begin{figure}[htbp]
\includegraphics[width=0.9\textwidth]{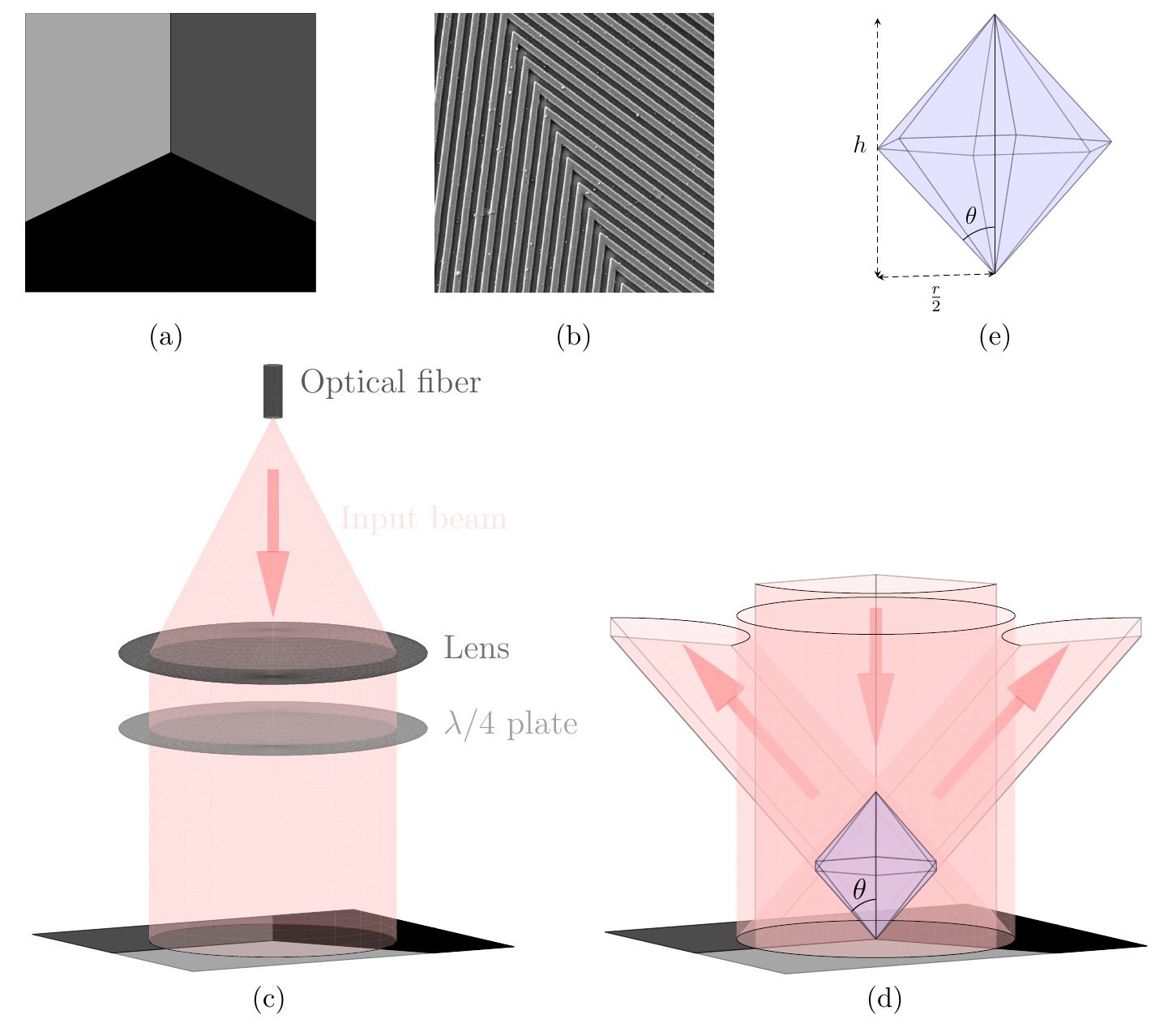}
\caption{\label{gmot}Schematic representation of MOT production using an optical grating. a) The grating chip consists of three identical linear gratings, lying in a plane with $120^{\circ}$ relative orientations. b) Scanning electron microscope (SEM) image showing the chip surface at a position where two gratings meet. c) A single linearly polarised input beam emanating from an optical fiber is collimated and subsequently made circularly polarised. d) The incoming and diffracted laser beams (red) together with the overlap volume (purple). e) The overlap volume approximately has the shape of a double hexagonal pyramid and has a rms height $h$. The incoming laser beam has a rms radius $r$ and is diffracted under an angle $\theta$.}
\end{figure}

The GMOT, developed at University of Strathclyde, Glasgow, is based on an optical grating that diffracts a single incoming circularly polarised laser beam. The MOT is formed inside the overlap volume which is spanned by the incoming beam, the zeroth order back reflection and three first order diffracted beams\cite{McGilligan2017,McGilligan2016,nshii2013,Cotter2016}. The grating chip consists of three identical linear gratings, lying in a plane with $120^{\circ}$ relative orientations, see Figure~\ref{gmot}a). Each grating diffracts the incoming laser beam according to Bragg's law,
\begin{equation}
n \lambda = d_{\rm g} \sin(\theta)
\end{equation} 
with $d_{\rm{g}}$ the grating period, ($n=\pm1$) the diffraction order and $\lambda=780~$nm the Rubidium trapping laser wavelength. We have used two gratings with periods $d_{\rm{g}}=892~\rm{nm}$ and $d_{\rm{g}}=1560~\rm{nm}$ which result in first order diffraction angles $\theta=61^{\circ}$ and $\theta=30^{\circ}$ respectively. The diffraction angle is defined as the angle between the grating normal and the diffracted beam, see Figure~\ref{gmot}d and e. Figure~\ref{gmot}b shows a scanning electron microscope (SEM) image of the $\theta=61^{\circ}$ grating structure at a position where two gratings meet. Higher diffraction orders ($n\geq 2$) are cut off because $d_{\rm g}<2\lambda$. The radial force balance\cite{McGilligan2016} for both gratings is automatic when the incoming beam is aligned with the center of the grating.


The $\theta=61^{\circ}$ chip measures $20~\times~20~\rm{mm}$ while the $\theta=30^{\circ}$ chip measures $28~\times~28~\rm{mm}$. Both chips are manufactured in silicon and have a $100~$nm top layer of aluminium which reflects the incoming laser beam. The gratings are fabricated using electron beam lithography and have a $50:50$ duty cycle and a $\lambda/4=195~$nm etch depth\cite{nshii2013}.


Figure~\ref{gmot}c shows a single linearly polarised laser beam emanating from an optical fiber, which is collimated using a lens and subsequently circularly polarised using a quarter wave plate. Figure~\ref{gmot}d shows that when the center of the beam is aligned with the center of the grating three first order diffracted beams are created which span the overlap volume (purple). The overlap volume approximately has the shape of a double hexagonal pyramid as is shown in Figure~\ref{gmot}e. The height of the overlap volume is $h=r\cot(\theta)$ with $r$ the rms radius of the laser beam. The experiments using the $\theta=61^{\circ}$ chip have been done with a Gaussian input laser beam with a $1/e^{2}$ beam radius of $r=7.5~\rm{mm}$. The experiments using the $\theta=30^{\circ}$ chip were done with a flattop input beam with a radius $r=12.5~\rm{mm}$. Using these values we calculate $h_{61}\approx 4~\rm{mm}$ and $h_{30}\approx 21~\rm{mm}$. The cold gas can be trapped anywhere inside the overlap volume. The exact position of the atom cloud is determined by the zero point of the magnetic quadrupole field. The ($1/e^{2}$) trap volume is given by

\begin{equation}
V=\frac{\sqrt{3}}{8}~r^{3}\cot(\theta).
\end{equation}

With the above mentioned beam parameters we calculate the overlap volume $V_{61}\approx 51~\rm{mm}^{3}$ and $V_{30}\approx 730~\rm{mm}^{3}$. The atom number $N$ scales\cite{Lindquist1992} with the overlap volume $V$ according to $N\propto V^{1.2}$. The Riis group have reported\cite{nshii2013} that it is possible to trap $2\cdot10^{7}$ atoms using a grating MOT with an overlap volume of $V\approx 570 ~\rm{mm}^{3}$. Using the scaling law we expect to be able to trap $N_{61}\approx10^{6}$ and $N_{30}\approx 3\cdot 10^{7}$ atoms in our experiment. 

\subsection{Ultracold electron source}

\begin{figure}[htbp]
\centering	
\includegraphics[width=13cm]{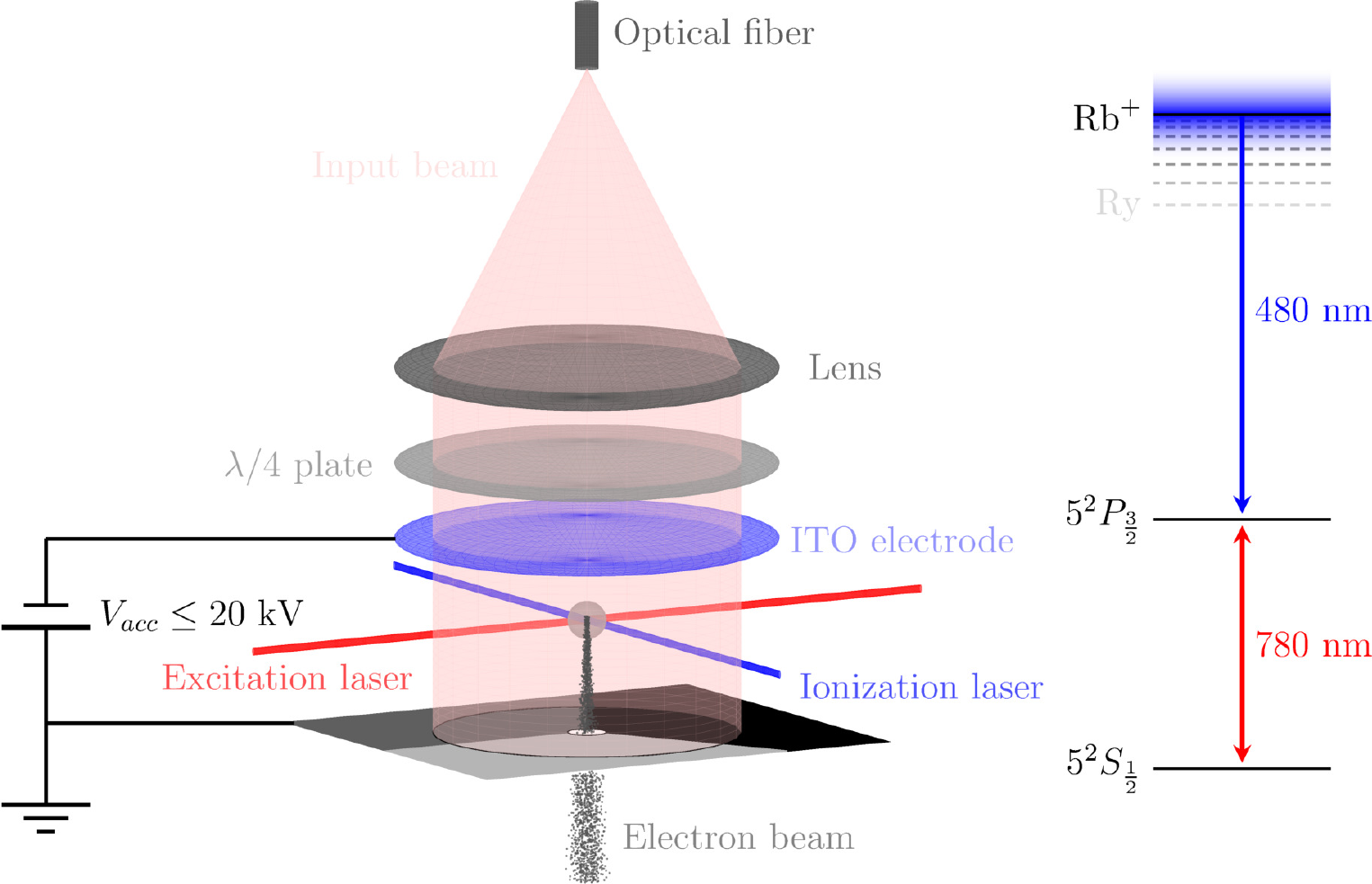}
\caption{Schematic representation of the ultracold electron source based on a GMOT. A two-step ionisation scheme is used. The excitation and ionization laser define a volume where electrons are created. This results in an approximately Gaussian spherical ionization volume with a rms radius $\sigma_{x}=30~\mu\rm{m}$. These electrons are accelerated towards the grounded grating and pass through a hole in the center.}
\label{gmotelectron}
\end{figure}

Figure~\ref{gmotelectron} shows a schematic representation of the electron gun. First the MOT is loaded with $85$-rubidium atoms, then the trapping beam is switched off for a few $\mu\rm{s}$ so that all atoms relax back to the ground state. During these few $\mu\rm{s}$ a small cylinder ($60~\mu\rm{m}$ waist) of atoms is excited using a cw excitation laser beam ($5^{2}S_{\frac{1}{2}}\rightarrow5^{2}P_{\frac{3}{2}}$). The excitation laser beam is intersected by a pulsed $480~\rm{nm}$ ionisation laser beam at right angles which is focused down to a $60~\mu\rm{m}$ waist. This results in an approximately Gaussian spherical ionization volume with a root-mean-square (rms) radius $\sigma_{x}=30~\mu\rm{m}$. The wavelength $\lambda_{\rm{ion}}$ of the ionisation laser is tuned close to the ionisation threshold to minimise the excess energy $E_{\rm{exc}}$ gained by the electron. The Stark shifted excess energy is given by

\begin{equation}
E_{\rm{exc}}= hc \left(\frac{1}{\lambda_{\rm{ion}}}-\frac{1}{\lambda_{0}}\right) + 2 E_{h} \sqrt{\frac{E}{E_{0}}},
\end{equation}

with $\lambda_{0}=479.06~\rm{nm}$ the zero-field ionization laser wavelength threshold, $E_{h}=27.2~\rm{eV}$ the Hartee energy, $E$ the applied electric field across the MOT, $E_{0}=5.14 \cdot 10^{11}~\rm{V}/\rm{m}$ the atomic unit of electric field, $h$ plank's constant and $c $ the speed of light. Ultracold electron bunches have been generated by using nanosecond ionization laser pulses~\cite{Claessens2007,Taban2010} and later by femtosecond pulses~\cite{Engelen2013,McCulloch2013}. The length of the ionization volume determines the energy spread\cite{franssen_pulse,Franssen_energy} of the electron beam. Additionally, the initial electron distribution can be tailored\cite{McCullochShaping} in 3D by shaping of the ionisation and excitation laser beams. 

To extract an electron beam a static electric field is applied across the trapped cloud of atoms. The negative electrode is an Indium-Tin-Oxide (ITO) coated quartz disk that is transparent for the trapping laser beam; the grounded grating chip is the positive electrode, which has a hole in the center, allowing the electron beam to pass through. The distance of the ionisation volume to the grating combined with the applied voltage determine the amount of energy the electron beam acquires.

\section{Detailed experimental Design}\label{design}

\begin{figure}[h!]
\centering	
\hspace{3cm}
\includegraphics[width=0.9\textwidth]{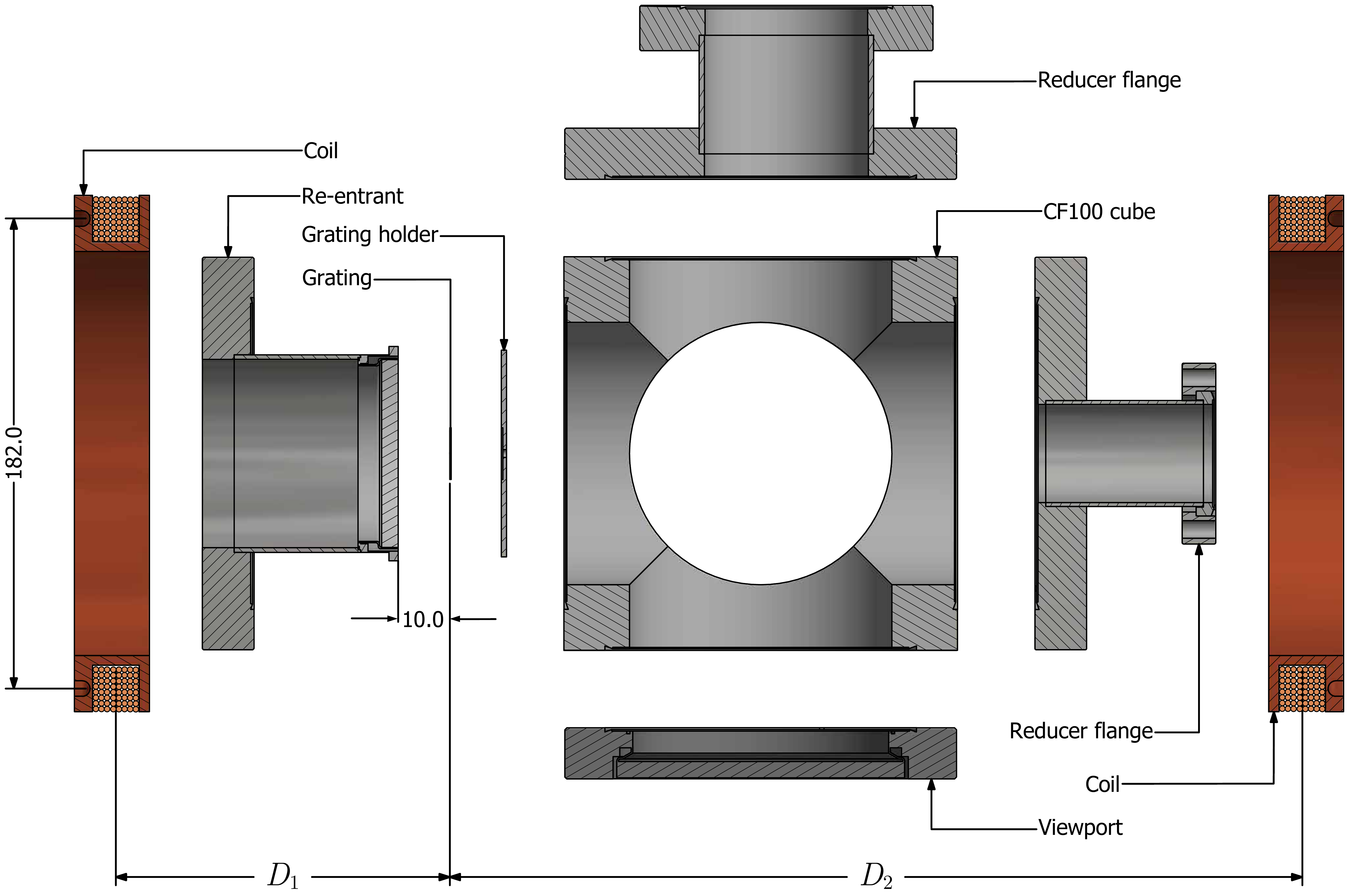}
\caption{Cross-section of the experimental setup. The grating is fixed in a holder which is connected to the re-entrant flange using $10~\rm{mm}$ spacers. The two coils generate the magnetic quadrupole field with $D_{1}$ and $D_{2}$ the distance from the coils to the grating surface.}
\label{vacuumsystem}
\end{figure}

We have designed a modular compact turn-key ultracold electron source that offers maximum optical accessibility. To keep the design simple we decided to have both the MOT coils and the high-voltage (HV) outside the vacuum, avoiding vacuum feed-throughs. A breakout cross-section of the vacuum system is depicted in Figure~\ref{vacuumsystem}. The main body of the electron gun consists of a CF100 cube. The left flange is a re-entrant flange which allows the accelerator module to be mounted close to the grating. The grating is embedded in a holder which is connected to the re-entrant flange. 

The right flange is a reducer flange which couples the cube to the electron beamline. The bottom, front and back of the cube are sealed with CF100 UV grade viewports. The top flange is a CF100 to CF63 reducer which houses the rubidium dispensers providing a rubidium background pressure of $\sim 2.5 \cdot 10^{-9}~$mbar. The base pressure of the vacuum system is $< 1 \cdot 10^{-9}~$mbar. The two coils generate the magnetic quadrupole field that is required to trap the rubidium atoms. 

\subsection{Quadrupole field}

The magnetic field coils are asymmetrically driven to compensate for the fact that the trapped gas cloud is not in the center of the CF100 cube ($D_{2}>D_{1}$), see Figure~\ref{vacuumsystem}. Due to the large coil radius and the large distance from the coils to the trapped gas we need $\sim 2500$ Ampere turns to provide the desired axial magnetic field gradient of $\sim 0.15~\rm{T/m}$. Both coils have $196$ turns and a radius $R_{\rm{coil}}=91~\rm{mm}$. The distance from the first coil to the grating surface is $D_{1}=79~\rm{mm}$ and the distance from the second coil to the grating surface $D_{2}=100~\rm{mm}$. The distance $D_{\rm{M}}$ between the gas cloud and the grating surface determines the ratio between the coil currents:

\begin{equation}
\frac{I_{1}}{I_{2}}=-\left(\frac{\left(D_{1}-D_{\rm{M}}\right)^{2}+R_{\rm{coil}}^{2}}{\left(D_{2}+D_{\rm{M}}\right)^{2}+R_{\rm{coil}}^{2}}\right)^{3/2}, \label{currentfraction}
\end{equation} 

with $I_{1}$ and $I_{2}$ the currents running through the first coil and second coil, respectively. Typically we operate the coils at $I_{1}=-10~$A and $I_{2}=15~$A. This results in an axial gradient $(\nabla B)_{z}=0.2~\rm{T/m}$. The position $D_{\rm{M}}$ of the gas cloud inside the overlap volume can be controlled by changing the current running through either one of the coils. Typically the MOT is formed in the center of the overlap volume which means that $D_{\rm{M}} \approx 2~\rm{mm}$ or $D_{\rm{M}} \approx 10~\rm{mm}$ depending on the grating that is used, see Section~\ref{MOTvol}.

\subsection{Accelerator}\label{accel}

\begin{figure}[htbp]
\centering	
\includegraphics[width=1\textwidth]{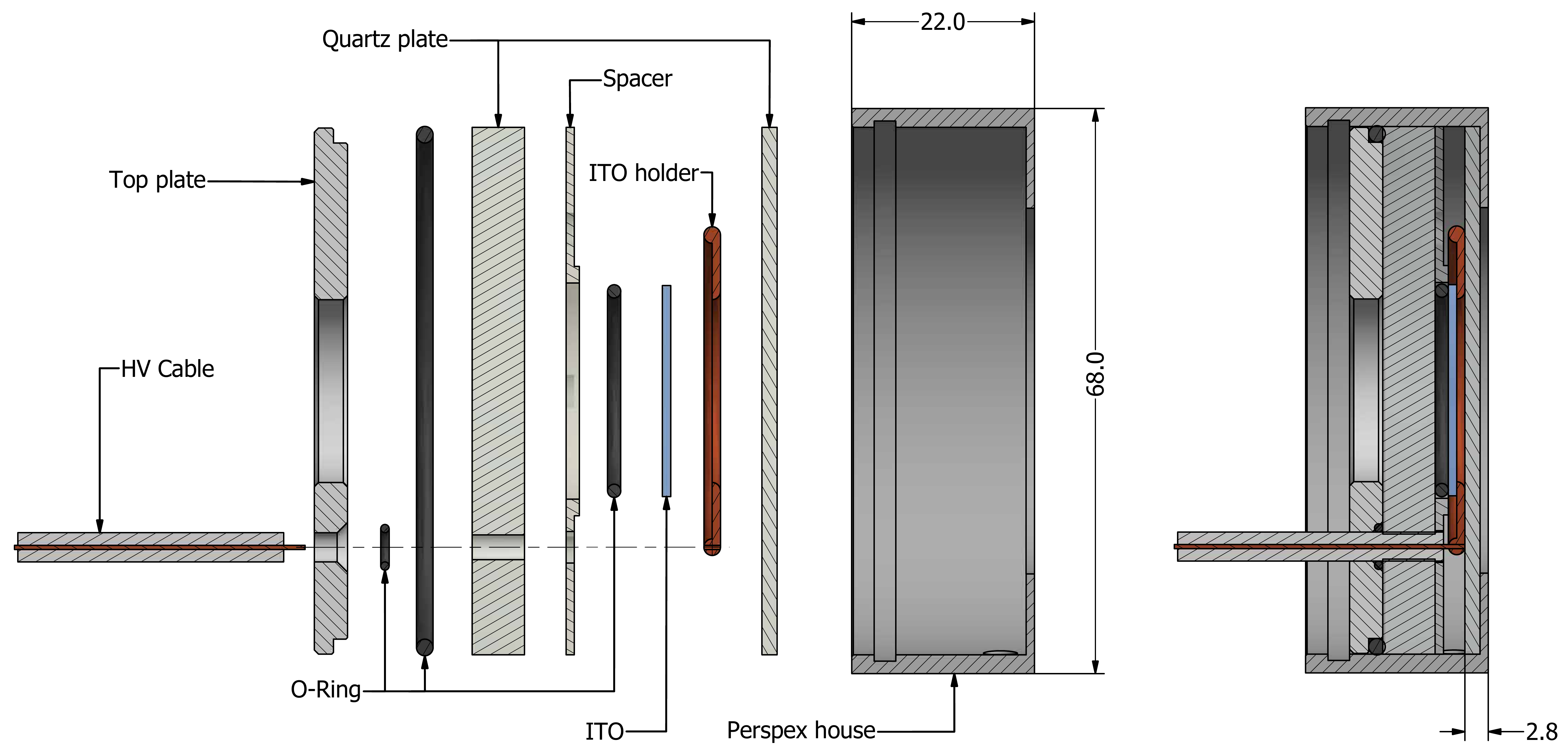}
\caption{Cross-sectional view of the accelerator module. Left: break out view of the accelerator module. Right: Assembled module. The transparent ITO electrode is connected to a copper electrode ring (ITO holder) using conductive epoxy. The copper ring is connected to a HV power supply. The electrode is sandwiched between two quartz plates.}
\label{accelerator}
\end{figure}

To reach electron energies of $\sim 10~\rm{keV}$ we need an electric field strength of $\sim 1~\rm{MV/m}$ across a $\sim 1~\rm{cm}$ gap. To avoid HV feedthroughs we designed a HV module which is mounted outside the vacuum, inspired by Reference\cite{Gunton}. The accelerator module can be inserted into the re-entrant flange, see Figure~\ref{vacuumsystem}. The electrode is an Indium Tin Oxide (ITO) coated quartz plate and therefore transparent for $780~$nm light. Figure~\ref{accelerator} shows a cross-sectional view of the accelerator module.

The accelerator house is made from perspex, the bottom layer is a quartz disk. On top of this disk we glued a copper ring (ITO holder) which is connected to a high voltage power supply. The ITO plate is glued onto the copper ring with conductive epoxy, electrically connecting the copper and the ITO surface. On top of the ITO plate we have an O-ring and a spacer holding it in place. On top of this we have another quartz plate and a top plate which is used to compress all the O-rings to make sure that the cavities are sealed. The volume in the assembly between the two quartz plates but outside the O-ring is filled with epoxy. In this way the HV electrode is shielded from the environment with a high dielectric constant material, while keeping the compartments where the trapping laser passes through free from epoxy.

\begin{figure}[htbp]
\centering	
\includegraphics[width=0.7\textwidth]{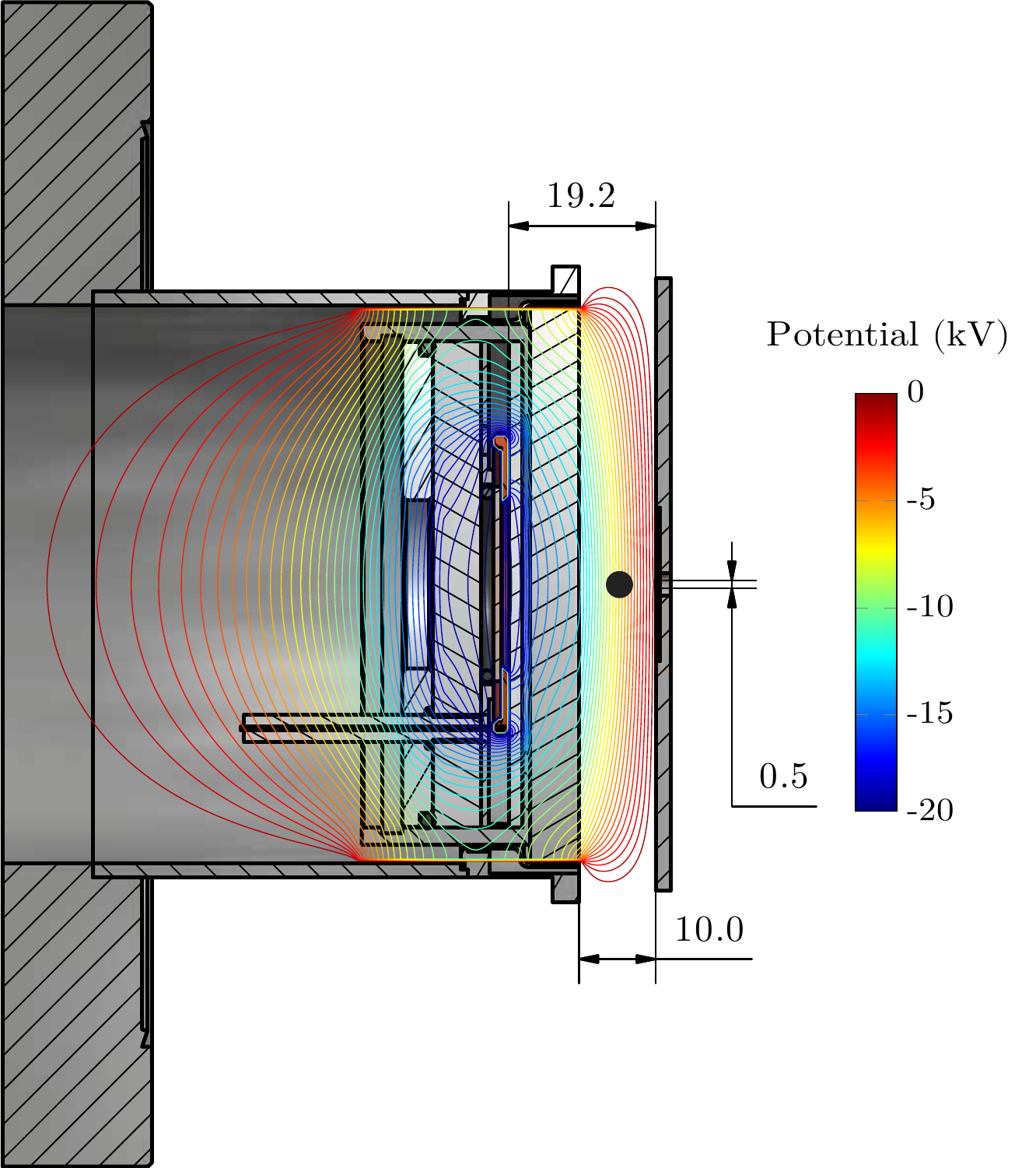}
\caption{Cross-sectional view of the accelerator module mounted in the re-entrant flange. The gas cloud (black dot) is created in the $10~\rm{mm}$ gap between the viewport and the grating. The figure shows the equipotential lines of the calculated electrostatic field generated by the transparent electrode. The electrode is at $V_{\rm{acc}}=-20~\rm{kV}$ and is separated $19.2~\rm{mm}$ from the grounded grating holder. The hole in the grating has a diameter of $0.5~\rm{mm}$.}
\label{accelerator_reentrant}
\end{figure}

\begin{figure}[htbp]
\centering	
\includegraphics[height=0.35\textwidth]{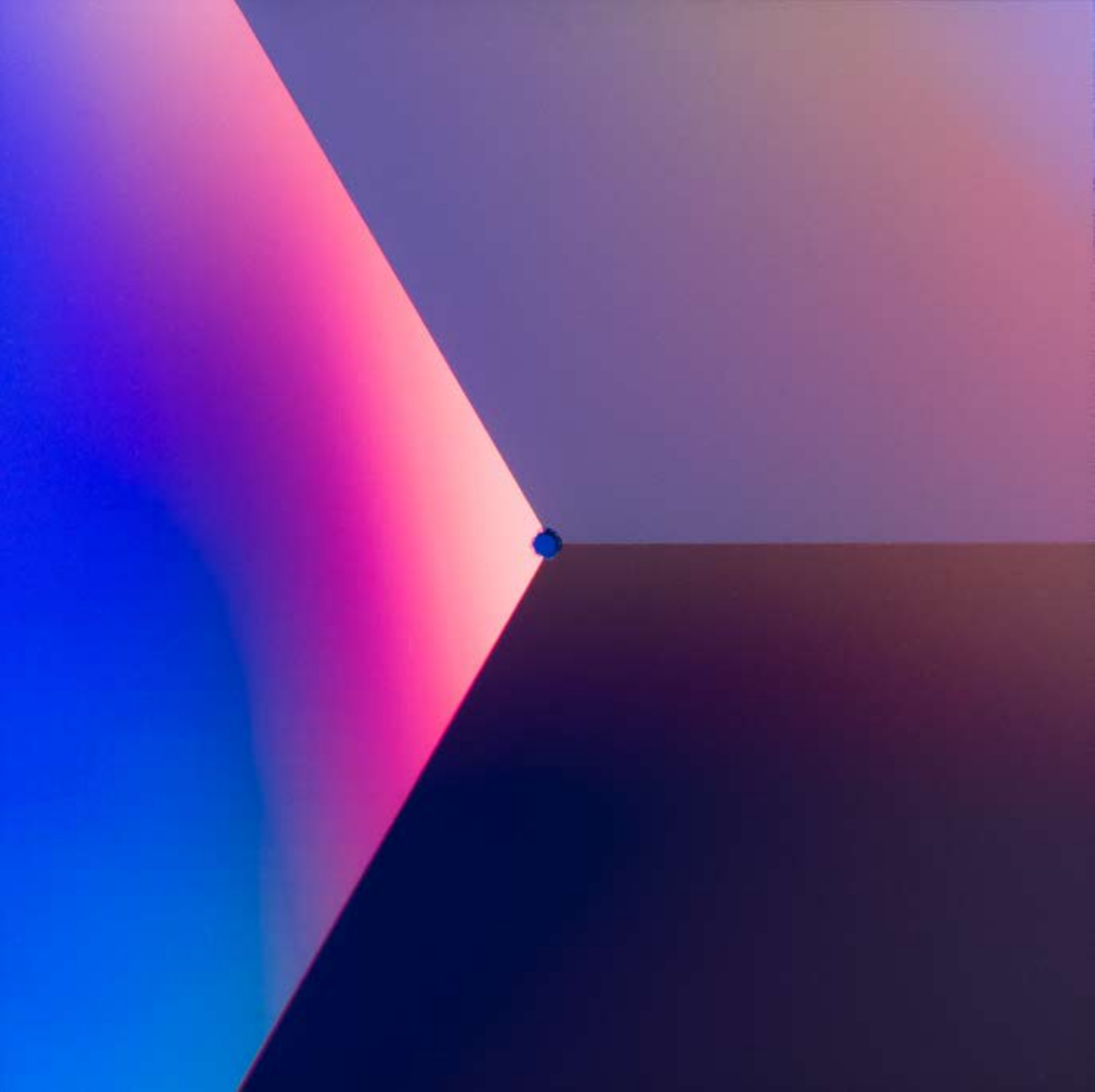}
\includegraphics[height=0.35\textwidth]{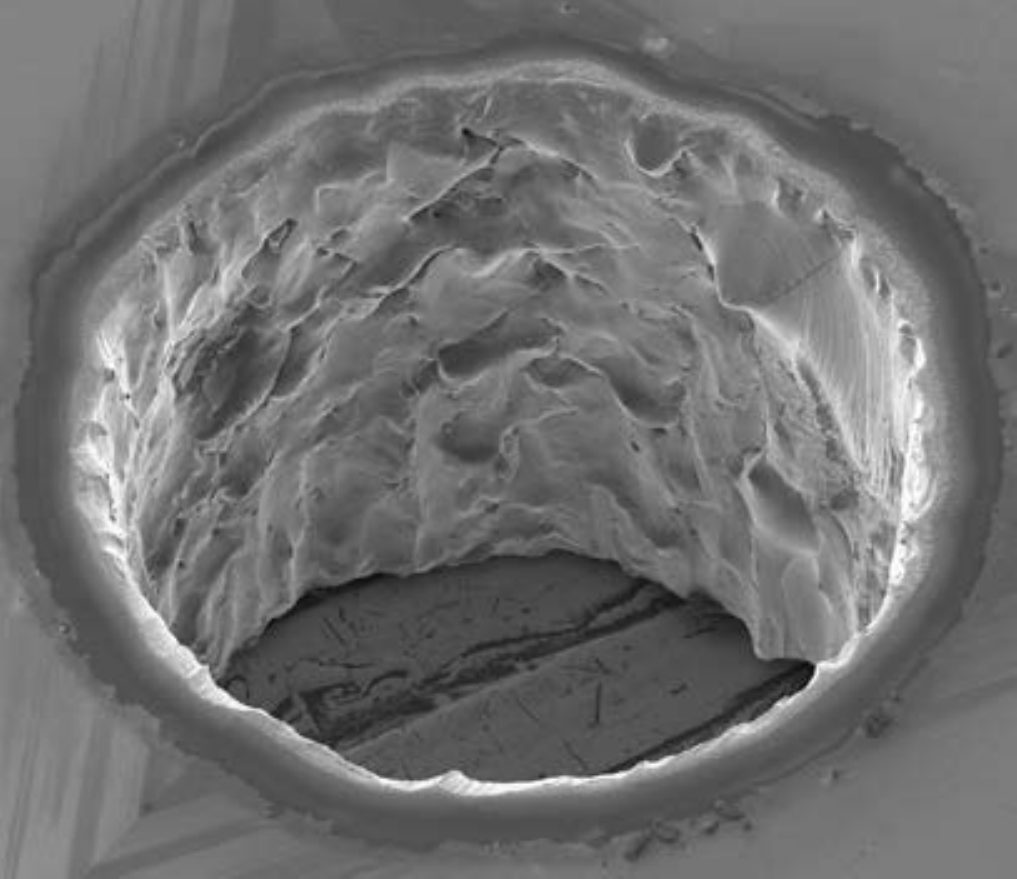}
\caption{Left: The grating chip with hole in the center. Right: SEM image (closeup) of the hole in the center.}
\label{GratingHoleDrilled}
\end{figure}

Figure~\ref{accelerator_reentrant} shows a cross-sectional view of the accelerator module mounted in the re-entrant flange, which is pushed against the viewport using a styrofoam spacer. The high voltage electrode and the grounded grating support are separated by $19.2~\rm{mm}$. The figure also shows the equipotential lines of the acceleration field with the electrode at $V_{\rm{acc}}=-20~\rm{kV}$. The gas cloud (indicated by the black dot) is trapped in between the viewport and the grating, which are separated by $10~\rm{mm}$ of vacuum. The accelerated electrons pass through a hole in the grating with a diameter of $0.5~\rm{mm}$. Figure~\ref{GratingHoleDrilled} shows an image of the grating with the hole in the center (left) and a closeup SEM image of the hole (right).

\begin{figure}[htbp]
\centering	
\includegraphics[width=0.9\textwidth]{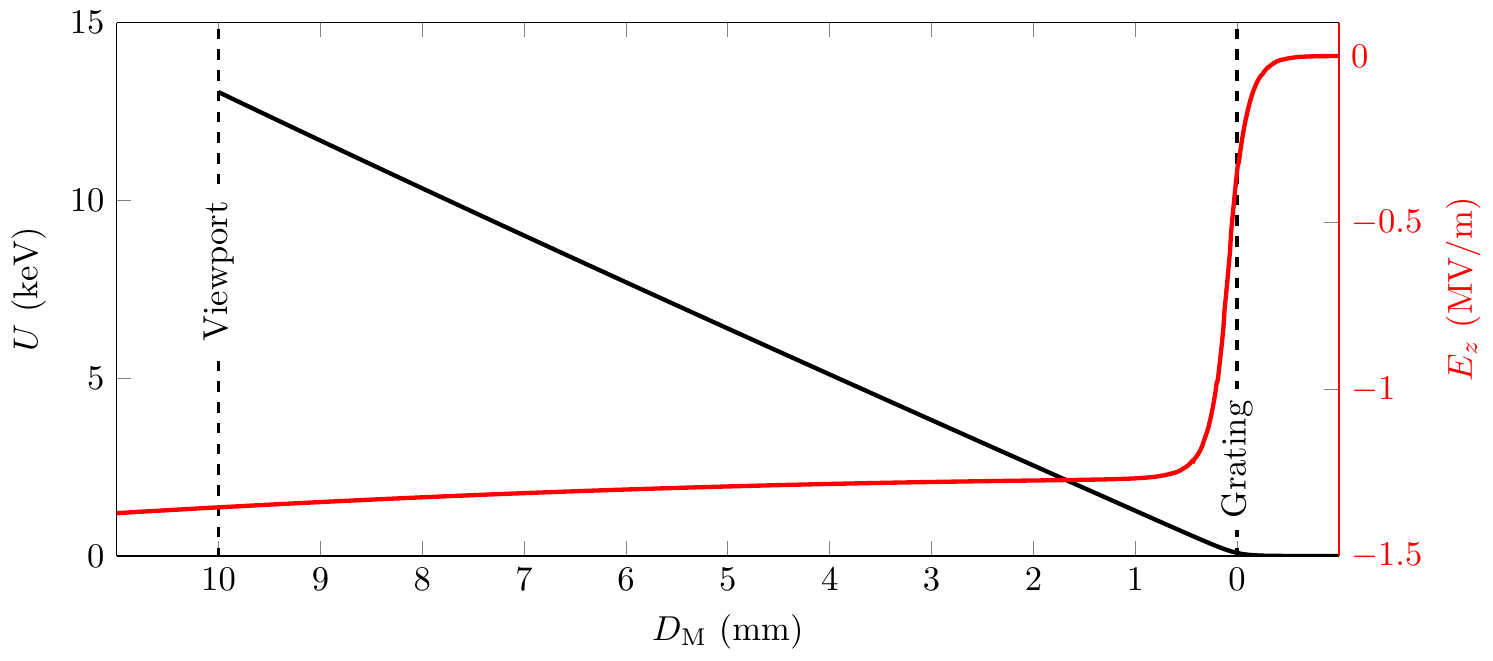}
\caption{Final kinetic energy $U$ (black) and the simulated electric field $E_{z}$ (red) as a function of distance $D_{\rm{M}}$ between the gas cloud and the grating.}
\label{accelekin}
\end{figure}

Figure~\ref{accelekin} shows the simulated electric field $E_{z}$ and the final electron kinetic energy $U$ reached as a function of distance $D_{\rm{M}}$ between the gas cloud and the grating for an accelerator potential $V_{acc}=-20~\rm{kV}$. The further away the electrons are created from the grating the larger the final kinetic energy $U$. Based on the height of the overlap volume we expect to create at MOT at $D_{\rm{M}} \approx 2~\rm{mm}$ or $D_{\rm{M}} \approx 10~\rm{mm}$ depending on the grating that is used, see Section~\ref{MOTvol}. This would result in an electron beam energy $U \approx 2.5~\rm{keV}$ for the $\theta=61^{\circ}$ chip and $U\approx 10~\rm{keV}$ for the $\theta=30^{\circ}$ chip. 

\subsection{Beamline}

\begin{figure}[htbp]
\centering	
\includegraphics[width=0.9\textwidth]{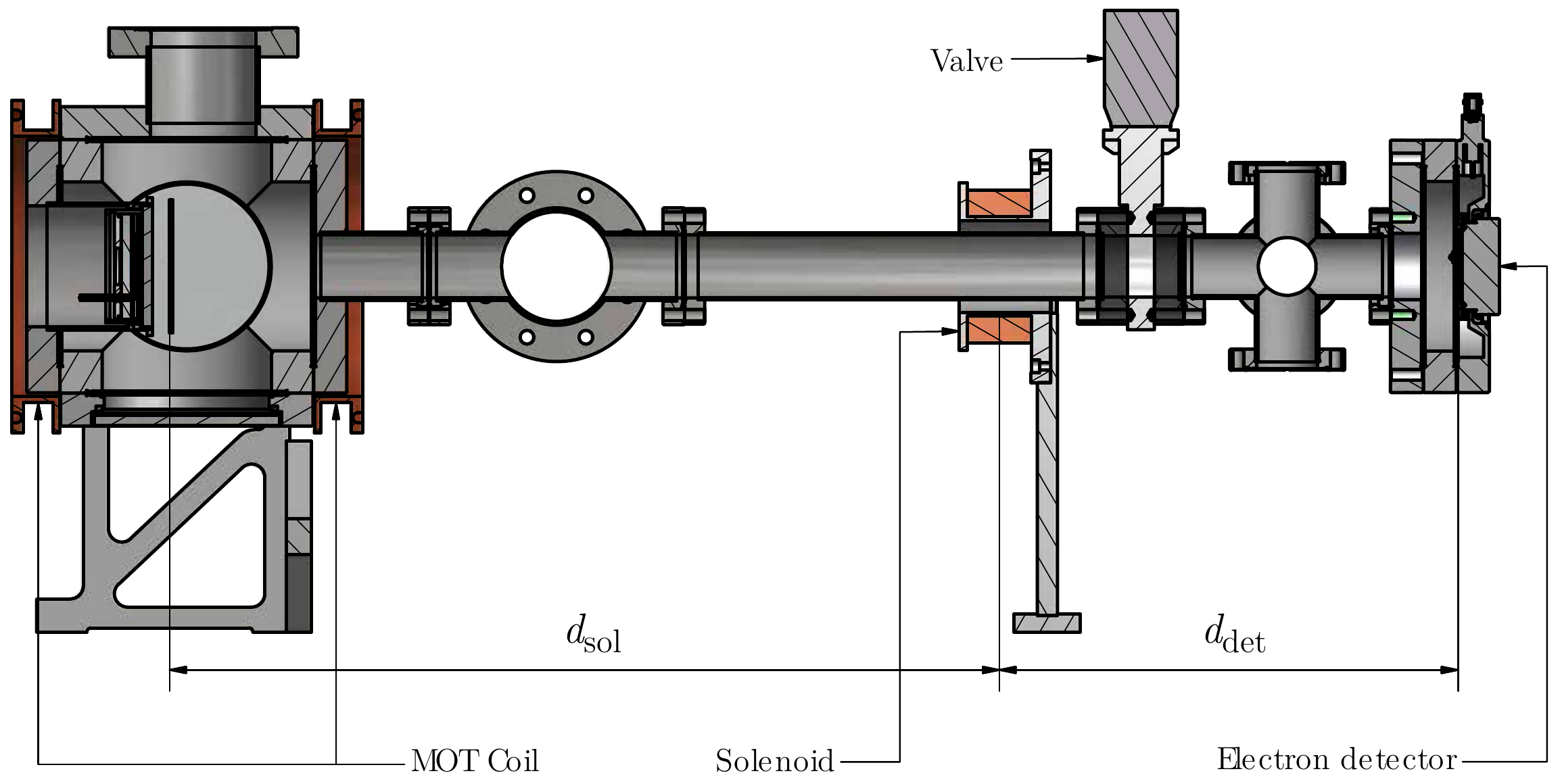}
\caption{Schematic representation of the electron gun and beamline.}
\label{electronbeamline}
\end{figure}

Figure~\ref{electronbeamline} shows a cross-sectional view of the electron gun and beamline. The electrons are created in the vacuum space between the re-entrant and the grating support, see Figure~\ref{accelerator_reentrant} for a more detailed view. The divergence and size of the electron beam can be controlled by the magnetic solenoid lens positioned at a distance $d_{\rm{sol}}=484~\rm{mm}$ from the gas cloud. The detector assembly is positioned $d_{\rm{det}}=293~\rm{mm}$ behind the magnetic solenoid lens. The detector assembly consists of a micro-channel plate (MCP) and a phosphor screen which is imaged onto a CCD camera.

\section{Commissioning Grating MOT}\label{sectionexperi}

We have tested the operation of the GMOT under the influence of an applied electric field and with a hole in the center of the grating chip. The measurements were done using the $\theta=61^{\circ}$ grating chip. The input laser beam has $P_{\rm{t}}=22.5~\rm{mW}$ at the trapping wavelength and $P_{r}=5.5~\rm{mW}$ at the repump wavelength. The trapping beam has a $1/e^{2}$ beam diameter of $15~\rm{mm}$ resulting in a trapping beam peak intensity $I_{\rm{t}}=25.5~\rm{mW}/\rm{cm}^{2}$, which is well above the rubidium saturation intensity~\cite{metcalf}. The repump beam peak intensity is $I_{\rm{r}}=6.2~\rm{mW}/\rm{cm}^{2}$. The atom numbers were estimated using fluorescence measurements with a saturated atomic scattering rate $\Gamma/2$.

Typically $4.2 \times 10^{6}$ atoms can be loaded in $95~\rm{ms}$, corresponding to a loading rate $4.4 \times 10^{7}~ \rm{s}^{-1}$. This implies that the loading rate is sufficient to extract electron bunches containing on the order of $10^{3}$ electrons per pulse at a repetition rate of $1~\rm{kHz}$. The peak atom density $n_{\rm{atom}}=2.1\cdot 10^{16}~\rm{m}^{-3}$ is reached for a magnetic field gradient $\nabla B_{z} = 0.2~\rm{T/m}$.  

\subsection{DC Stark shift}

The detuning at which the number of atoms in the trap is maximal changes if an electric field is applied due to Stark shifting of the cooling transition frequency. The energy levels for the hyperfine states of the $5^{2}S_{1/2}$ ground state $85$-rubidium atom all shift equally~\cite{Krenn1997}. The excited state $5^{2}P_{3/2}$ energy levels are shifted differently depending on the hyperfine state due to a non-zero tensor polarisability\cite{Krenn1997}. 

\begin{figure}[htbp]
\centering	
\includegraphics[width=0.9\textwidth]{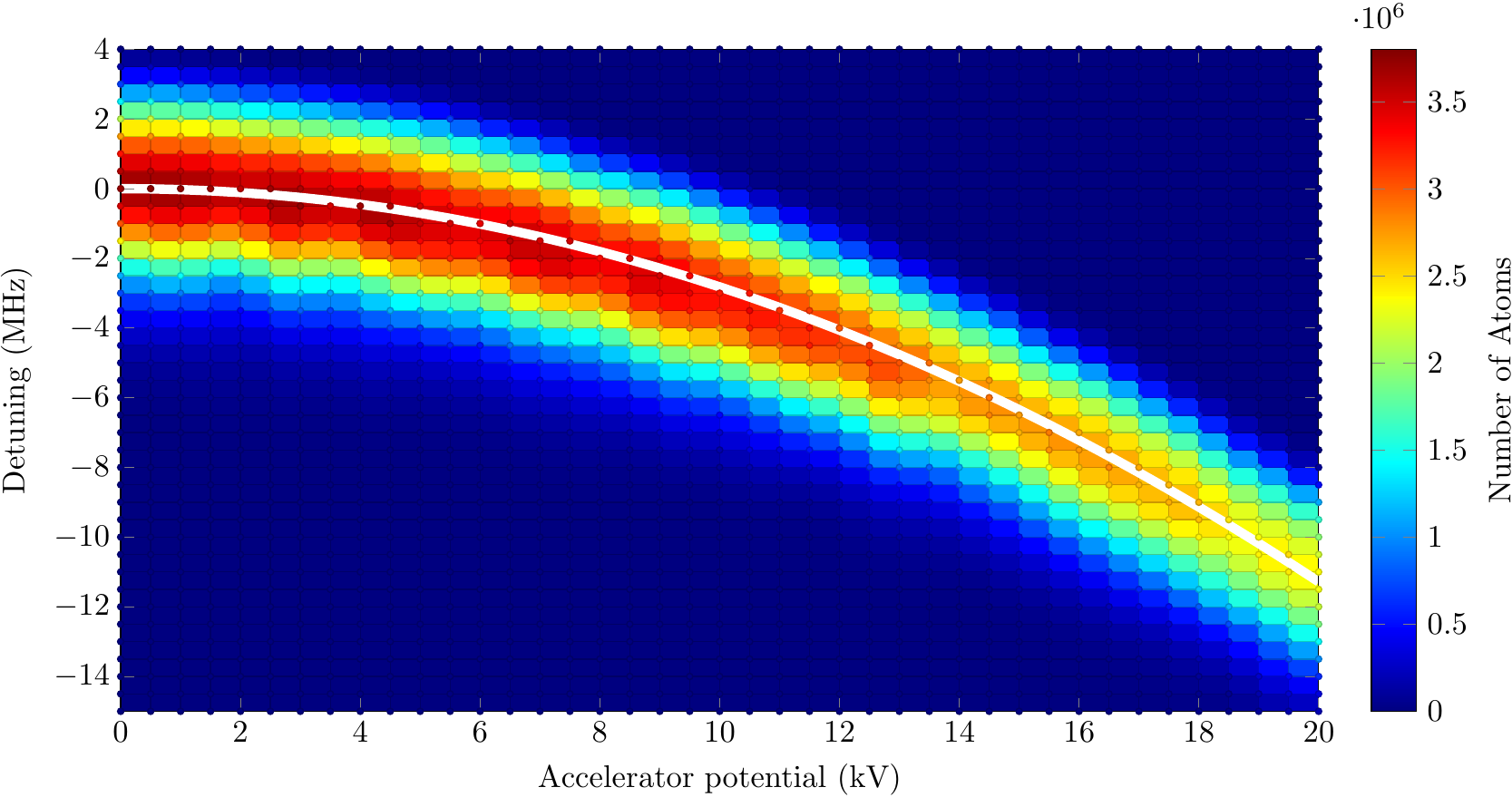}
\caption{Number of atoms in the MOT as a function of both accelerator potential and detuning with respect to the optimal detuning ($\delta_{E=0}=-9~\rm{MHz}$) when no electric field ($E=0$) is applied. The white line is a fit with theory and results in $E=1.2~\rm{MV/m}$.}
\label{Stark_shift_no_hole}
\end{figure}

Figure~\ref{Stark_shift_no_hole} shows the number of atoms in the trap as a function of accelerator potential and detuning with respect to the optimal detuning $\delta_{E=0}=-9~\rm{MHz}$ when no electric field ($E=0$) is applied. The white line shows the calculated Stark shift, for the transition $F=3$ to $F^{\prime}=4$ for final $m_{F^\prime}=0$. We assumed the Stark shift to be small compared to the hyperfine splittings\cite{Krenn1997} which holds for $E\leq 2~\rm{MV/m}$. The white curve corresponds to an electric field $E=1.2~\rm{MV/m}$ at the position of the gas cloud. This agrees well with the simulated electric field strength at a position $D_{\rm{M}}\approx 2~\rm{mm}$ for an accelerator potential of $V_{\rm{acc}}=-20~\rm{kV}$, shown in Figure~\ref{accelekin}.

Figure~\ref{Stark_shift_no_hole} also shows that the number of atoms in the trap decreases as the electric field is increased. This is due to broadening of the laser cooling transition by breaking of the degeneracy of the $m_{F}$ levels in the $5^{2}P_{3/2}$ state\cite{Gunton}. At maximum electric field we still have a steady state atom number $N_{\infty}=2.4\cdot 10^{6}$ which is more than sufficient to operate the GMOT as an electron source.

\section{Commissioning UCES}\label{ucescomm}

In the previous section we showed that $\sim 10^{6}$ atoms are trapped in the presence of a $\sim1~\rm{MV/m}$ acceleration field using the $\theta=61^{\circ}$ grating chip. The MOT was ionised using a tuneable nanosecond ionisation laser pulse with a rms pulse duration of $\sim 2.5~\rm{ns}$. The electron beam energy was measured by an electron time-of-flight scan\cite{Engelen2014} which resulted in a modest electron beam energy $U\approx 2~\rm{keV}$. This is also what we expect on the basis of the position of the gas cloud $D_{\rm{M}}\approx 2~\rm{mm}$ for the $\theta=61^{\circ}$ grating chip, see Section~\ref{accel}.
We decided to replace the $\theta=61^{\circ}$ diffraction grating by a $\theta=30^{\circ}$ grating which allowed us to create the gas cloud  $D_{\rm{M}} \approx 8~\rm{mm}$ above the grating surface, as expected from the height of the overlap volume. Additionally, this also increases the overlap volume which results in a gas cloud containing $\sim 10^{7}$ atoms. The larger value of $D_{M}$ increased the maximum achievable electron beam energy to $U\approx 10~\rm{keV}$. The input laser beam has a circular (diameter of $22~\rm{mm}$) flattop intensity distribution with $I_{t}\approx 15~\rm{mW}/\rm{cm}^2$ and $I_{r}\approx 3~\rm{mW}/\rm{cm}^2$.

In this section we present measurements of the electron beam energy, using a time-of-flight method, and of the transverse beam quality using a waist scan\cite{Engelen2013,Engelen2014} method. 

\subsection{Time-of-Flight}

\begin{figure}[htbp]
\centering	
\includegraphics[width=1\textwidth]{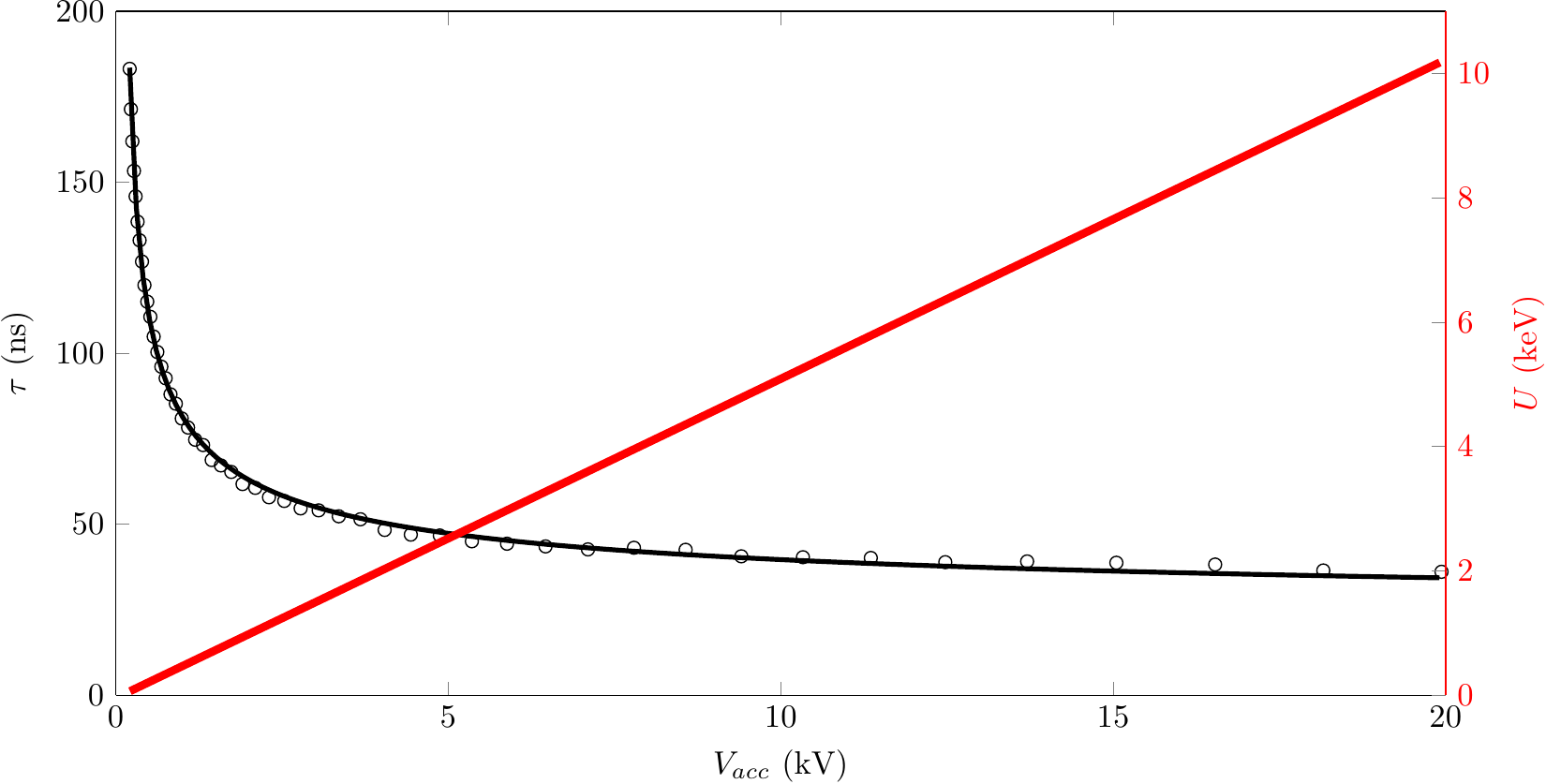}
\caption{Electron time-of-flight $\tau$ as a function of accelerator voltage $V_{\rm{acc}}$. The solid line is a fit using equation~\ref{electrontoffit}. The plot also shows the final electron beam energy $U$ as a function of accelerator voltage which was calculated using Equation~\ref{energyU} with $f$ and $V_{0}$ the fitted parameters. }
\label{electrontof}
\end{figure}

The electron beam energy was determined using an electron time-of-flight (TOF) scan\cite{Engelen2014}. The arrival time of the electron pulse on the MCP was determined by measuring the charge signal with a trans-impedance amplifier. 
Figure~\ref{electrontof} shows the arrival time of the electron pulse relative to the ionisation time as a function of acceleration voltage $V_{\rm{acc}}$. 

The resulting TOF data was fit, solid line in Figure~\ref{electrontof}, using the relativistically correct expression

\begin{equation}
\tau = \frac{d_{\rm{sol}}+d_{\rm{det}}}{c\sqrt{1-\left(\frac{mc^2}{mc^2 + f(V_{\rm{acc}}-V_{0})}\right)^{2}}}+\tau_{0}
\label{electrontoffit}
\end{equation}
with $c$ the speed of light, $m$ the electron mass, $V_{0}$ an accelerator potential offset and $\tau_{0}$ an electronics delay. This function was fitted with $f$, $V_{0}$ and $\tau_{0}$ as fitting parameters, yielding $f=0.51\pm 0.02$, $V_{0}=83\pm6 \rm{V}$ and $\tau_{0}=2.1\pm0.1~\rm{ns}$. The electron energy $U$ is given by

\begin{equation}
U=-ef(V_{\rm{acc}}-V_{0}),\label{energyU}
\end{equation}
with $e$ the electron charge. The final electron beam energy $U$ as a function of accelerator voltage $V_{\rm{acc}}$ calculated using Equation~\ref{energyU} is shown in Figure~\ref{electrontof}. The maximum achievable electron energy is $U=10.2\pm 0.4~\rm{keV}$ for an accelerator potential $V_{\rm{acc}}=-20~\rm{kV}$.

During operation, the re-entrant viewport gets coated with a thin layer of rubidium which reduces the work function of the quartz surface. Scattered ionisation laser photons give rise to photoemission which charges the glass plate, effectively creating a voltage offset $V_{0}$. Carefully aligning the ionisation laser and taking measure to prevent scattering of the laser light minimises the charging. Alternatively the viewport can be heated.

Additionally, the rubidium ions are accelerated towards the quartz plate. These ions can produce electrons by ion impact ionisation if they acquire sufficient kinetic energy. This causes a second electron signal on the electron detector which arrives about $150~\rm{ns}$ later than the electrons extracted from the Rubidium atoms. Switching the accelerator voltage off when the electrons have passed the grating should reduce the ion kinetic energy and thus the secondary electron yield. Alternatively, immediately switching the polarity of the field after the electrons have passed the grating will further decelerate the ions.

\subsection{Beam quality}

The transverse electron beam quality is given by the normalized rms transverse emittance which is defined by

\begin{equation}
\epsilon=\frac{1}{mc} \sqrt{\left<x^2\right>\left<p_{x}^2\right>-\left<xp_{x}\right>^{2}}	,
\end{equation}
 with $x$ the transverse position coordinate with respect to the bunch center and $p_{x}$ the transverse momentum with respect to the average transverse bunch momentum. The brackets $<...>$ denote averaging over the electrons in the bunch. The beam emittance is fixed at the source. Here the position and momentum coordinate are uncorrelated, i.e. $\left<xp_{x}\right>=0$, and therefore the emittance is equal to the product of beam size $\sigma_{x}=\sqrt{\left<x^2\right>}$ and uncorrelated transverse momentum spread $\sigma_{p_{x}}= \sqrt{\left<p_{x}^2\right>}$. The transverse size of the electron beam at the source was determined to be $\sigma_{\rm{x}}=30~\mu\rm{m}$ by measuring the spot size of the excitation laser beam in a virtual source point. The normalized beam emittance $\epsilon$ is related to the source temperature $T$ by 
\begin{equation}
\epsilon = \sigma_{\rm{x}}\sqrt{\frac{k_{b} T}{mc^{2}}}
\end{equation}
where we used $\sigma_{p_{x}}=\sqrt{m k_{b}T}$ with $k_{b}$ the Boltzmann constant. The normalised emittance $\epsilon$ can be determined from a waistscan\cite{Engelen2013,Engelen2014}. In a waist scan the size of the electron beam on the detector is measured as a function of focusing power of a magnetic solenoid lens. The ionisation laser wavelength was tuned to $\lambda_{\rm{ion}}=488.3\pm 0.1~\rm{nm}$ so that the electrons only receive a small amount of excess energy. At an electric field strength $F=0.37\pm0.04~\rm{MV}/\rm{m}$ this results in $-2.8\pm 3.4~\rm{meV}$ which should result in ultracold electron bunches\cite{Engelen2013,Engelen2014,McCulloch2013}.

\begin{figure}[htbp]
\centering	
\includegraphics[width=0.95\textwidth]{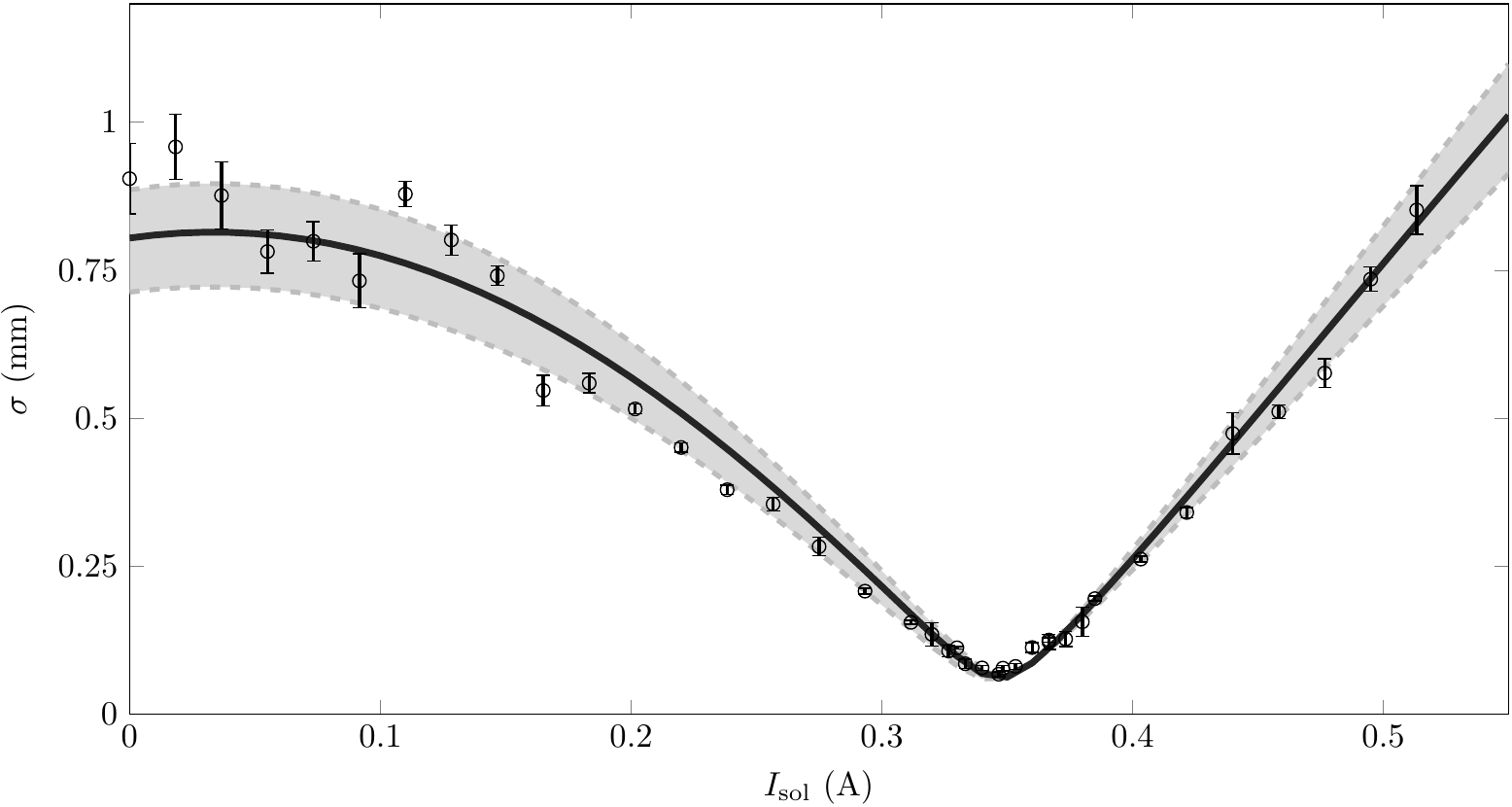}
\caption{The electron beam rms spot size $\sigma$ as measured on the detector as a function of current $I_{\rm{sol}}$ running through the magnetic solenoid lens.}
\label{electronwscan}
\end{figure}

Figure~\ref{electronwscan} shows the rms transverse beam size as measured on the detector as a function of current $I_{\rm{sol}}$ running through the solenoid. At the focus an rms electron spot size $\sigma \approx 50~\mu\rm{m}$ was reached which is at the limit of our detector resolution. Detailed charged particle tracking simulations were done, in which the source temperature was varied while the source size $\sigma_{x}$ was kept fixed. We used realistic fields for the accelerating field, the quadrupole magnetic field of the MOT coils and the field produced by the magnetic solenoid lens. Coulomb interactions can be neglected\cite{franssen_pulse} because we have used a nanosecond ionization laser pulse. The measured data was fitted with the simulation results\cite{Engelen2014} which was quadratically added to the detector resolution.  The solid black line in figure~\ref{electronwscan} is the best fit with particle tracing simulations.

The fit results in a beam emittance $\epsilon = 1.9~\rm{nm}\cdot\rm{rad}$, which corresponds to a source temperature $T=25~\rm{K}$. The grey band in the figure defines an upper and a lower limit (dashed lines) for the source emittance. At the lower limit this results in $\epsilon = 0.4~\rm{nm}\cdot\rm{rad}$ and at the upper limit $\epsilon = 2.8~\rm{nm}\cdot\rm{rad}$. Using $\sigma_{x}=30~\mu\rm{m}$, this translates into a lower temperature limit $T_{-}=1~\rm{K}$ and an upper temperature limit $T_{+}=50~\rm{K}$. The inner data points (close to the focus) imply a source temperature even lower than $25~\rm{K}$. The measured source temperature and beam emittance of the GMOT UCES are in line with results reported in earlier work~\cite{Engelen2013,Engelen2014,McCulloch2013}.


\section{Conclusion and outlook}\label{outlookconc}

We have successfully developed a compact ultracold electron source based on a GMOT. The unique modular design only requires one input beam that passes through a transparent accelerator module. We show that the GMOT can be operated with a hole in the center of the grating and with large electric fields applied across the trapping region. The electric field was determined by measuring the Stark shifts of the laser cooling transition. 

The electron beams extracted from the GMOT have been characterised. Beam energies up to $10\pm0.4~\rm{keV}$ were measured using a time-of-flight method. The normalised root-mean-squared transverse beam emittance was determined using a waist scan method, resulting in $\epsilon=0.4-2.8~\rm{nm}\cdot\rm{rad}$. Since the root-mean-squared transverse size of the ionisation volume is $30~\mu\rm{m}$ or larger, this implies an electron source temperature in the $1-50~\rm{K}$ range.

We have demonstrated a clear path towards harnessing the great potential of the UCES in a practical setting. Future research will focus on increasing the bunch charge of picosecond electron pulses created by femtosecond photoionisation. These pulses are sufficiently short for RF acceleration and compression, creating intriguing new possibilities. Obviously space charge forces will become a problem at higher bunch charges which can be addressed by shaping of the initial electron distribution.


\begin{acknowledgments}

We acknowledge valuable discussions with A. Arnold, E. Riis, P. Griffin, J. McGilligan and J. Conway. The grating was supplied by the Strathclyde grating MOT team. The hole in the grating was drilled by Tiny den Dekker from Philips innovation services.
This research is supported by the Institute of Complex Molecular Systems (ICMS) at Eindhoven University of Technology. The authors would like to thank E. Rietman and H. van Doorn for expert technical assistance.

\end{acknowledgments}

\bibliography{library.bib}

\end{document}